\documentclass[12pt,aps,prb,preprint]{revtex4}
\pdfoutput=1
\usepackage{graphicx}

\usepackage{graphicx}
\usepackage{rotating}
\usepackage{amsmath} 
\usepackage{amssymb}
\usepackage{mathptmx}
\bibpunct{}{}{,}{s}{}{,}

\begin{document}

\title{Advanced code-division multiplexers for superconducting detector arrays}
\thanks{Contribution of an agency of the U.S. government; not subject to copyright}

\author{K. D. Irwin}
\email{irwin@nist.gov}
\author{H. M. Cho}
\author{W. B. Doriese}
\author{J. W. Fowler}
\author{G. C. Hilton}
\author{M. D. Niemack}
\author{C. D. Reintsema}
\author{D. R. Schmidt}
\author{J. N. Ullom}
\author{L. R. Vale}
 \affiliation{National Institute of Standards and Technology, Boulder, CO 80305}

\date{\today}




\begin{abstract}

Multiplexers based on the modulation of superconducting quantum interference devices are now regularly used in multi-kilopixel arrays of superconducting detectors for astrophysics, cosmology, and materials analysis. Over the next decade, much larger arrays will be needed. These larger arrays require new modulation techniques and compact multiplexer elements that fit within each pixel. We present a new in-focal-plane code-division multiplexer that provides multiplexing elements with the required scalability. This code-division multiplexer uses compact lithographic modulation elements that simultaneously multiplex both signal outputs and superconducting transition-edge sensor (TES) detector bias voltages. It eliminates the shunt resistor used to voltage bias TES detectors, greatly reduces power dissipation, allows different dc bias voltages for each TES, and makes all elements sufficiently compact to fit inside the detector pixel area. These in-focal-plane code-division multiplexers can be combined with multi-gigahertz readout based on superconducting microresonators to scale to even larger arrays.

\end{abstract}

\maketitle

\section{Introduction}

Arrays of superconducting transition-edge sensors\cite{Irwin1995} (TES) are widely used to detect millimeter-wave, submillimeter, and x-ray signals. The development of kilopixel arrays has required cryogenic signal multiplexing techniques. To date, all deployed arrays use either time-division multiplexing (TDM)\cite{Chervenak1999} or frequency-division multiplexing (FDM)\cite{Yoon2001}. The potential advantages of multiplexing TES devices with Walsh codes have been anticipated\cite{Karasik2001,Podt2002}, and code-division multiplexing (CDM) circuits are now emerging\cite{Irwin2010} that can significantly increase the number of pixels multiplexed in each output channel, with more compact modulation elements.

Code-division multiplexing (CDM) shares many of the advantages of both TDM and FDM. In CDM, the signals from all TESs are summed with different Walsh-matrix polarity patterns. In the simplest case of two-channel CDM, the sum of the signals from TESs 1 and 2 would first be measured, followed by their difference. The original signals can be reconstructed from the reverse process. CDM can use the same room-temperature electronics as TDM. Unlike TDM, CDM does not suffer from the aliasing of SQUID noise by $\sqrt{N}$, where $N$ is the number of pixels multiplexed. CDM uses smaller filter elements and simpler room-temperature electronics than FDM, and it allows dc biasing of the TES sensor, making it easier to achieve optimal energy resolution\cite{Gottardi2009}.

Several CDM circuits have been demonstrated. One implementation, flux-summed CDM\cite{Irwin2010} ($\rm{\Phi}$-CDM), has been used to achieve average multiplexed energy resolution of 2.78 eV $\pm$ 0.04 eV FWHM at 6 keV with a small array of TES x-ray microcalorimeters\cite{Fowler2011}. 
Here we present a more advanced CDM multiplexing circuit topology that allows scaling to much larger multiplexing factors than $\rm{\Phi}$-CDM. In current-summed CDM (I-CDM), only one SQUID amplifier is required for each column of detectors. The current from all TES calorimeters in a column flows out in one pair of wires, with a coupling polarity that is switched at each pixel by compact, ultra-low-power switches in the focal plane itself. These switches can be placed underneath an overhanging x-ray absorber, so that separate wires need not be extracted from each pixel. In I-CDM, the voltage bias source for the TES calorimeters does not dissipate power at the cold stage, making the power dissipation in even megapixel arrays manageable. Because the dc voltage bias source for the TES calorimeters is naturally multiplexed, different bias voltages can be chosen for each pixel. Finally, the number of address wires required scales only logarithmically with the number of rows multiplexed. Logarithmic encoding of the address lines is made possible by the periodic nature of the response of the superconducting interferometer switches to address flux\cite{Irwin2010}.
\section{Superconducting polarity modulation switches}\label{switches}

\begin{figure}
\begin{center}
\includegraphics[%
  width=1.0\linewidth,
  keepaspectratio]{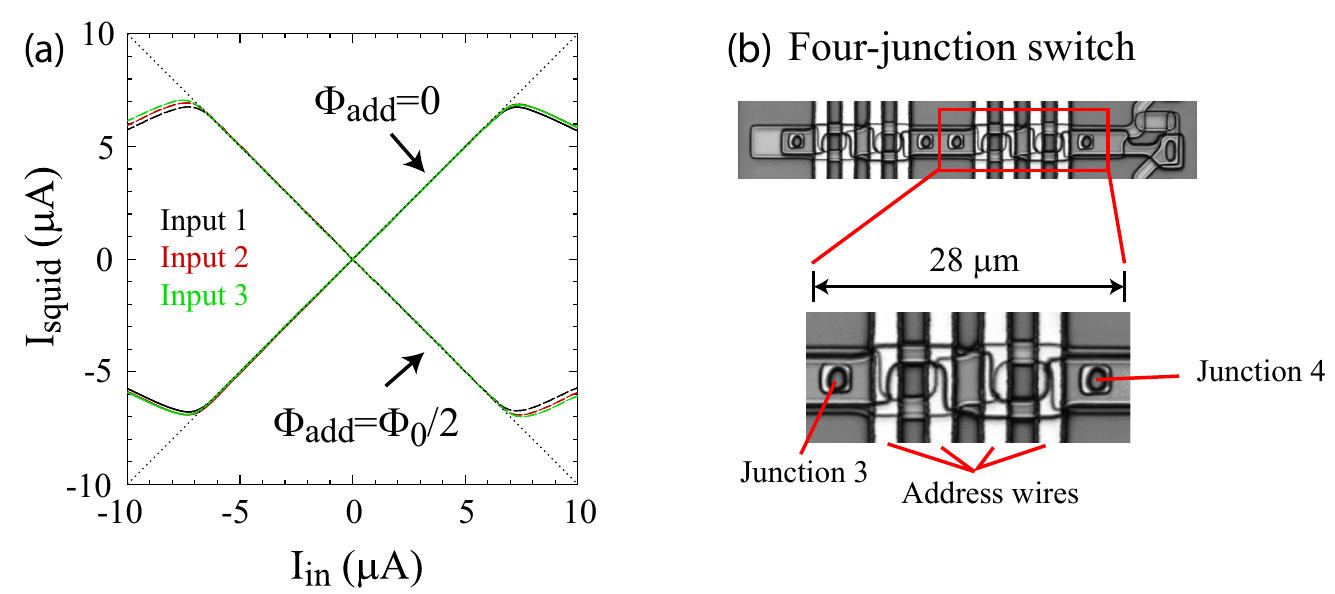}
\end{center}
\caption{
Polarity modulation. (a) Experimental measurements of a polarity modulator. Current into the SQUID ($I_{\rm{squid}}$) vs. the input current ($I_{\rm_{in}}$) in two different states: positive (solid, positive slope) data are for no applied address flux; negative (dashed, negative slope) data are for address flux $\Phi_{\rm{add}}=\Phi_0/2$. Data are shown for three different pixels summed into one SQUID (black, red, and green --- the data for all three curves are nearly identical). The inflection points near $\pm 7 \rm{\mu A}$ are indicative of the transition to the normal state above the current-carrying capacity of the modulator used in this experiment. 
(b) A photograph of the new generation of polarity-modulation switches that are used in I-CDM. The switch contains four Josephson junctions. Junction 1 and 2 are separated by a serial gradiometer, as are junctions 3 and 4. Junctions 2 and 3 are adjacent, and behave as a single junction with twice the critical current. This design provides larger operating margins and higher current-carrying capacity. An expanded view is shown for the serial gradiometer separating junctions 3 and 4.}
\label{fig:CurrentSteering}
\end{figure}

I-CDM requires a circuit that can modulate the polarity of the current coupling from a TES to its SQUID amplifier. The modulation has unity gain; amplification occurs only after the signal from many TES pixels is summed with different polarities. After the signal bandwidth is limited by a one-pole $L/R$ low-pass filter formed by a Nyquist inductor $L_{\rm{nyq}}$ and the TES resistance, superconducting switches steer the current from the TES into one of two pathways that couple to the SQUID with opposite polarity. Because the polarity modulation occurs at much higher frequency than the bandwidth of the signal, there is no degradation in performance from detector noise aliasing. We have already fabricated and tested appopriate polarity modulators\cite{Niemack2010} as part of a previous CDM circuit implementation. Fig. \ref{fig:CurrentSteering}a shows the current coupled to the SQUID ($I_{\rm{squid}}$) vs. the input current ($I_{\rm{in}}$) for the two settings of the modulator.

The polarity modulator contains superconducting switches\cite{Zappe1977, Beyer2008} that are based on compact, low-inductance SQUIDs controlled by an applied flux. These switches are designed so that their critical currents modulate from a maximum value at zero flux (zero applied address current) to very near to zero at a flux of $\Phi_0/2$. 
At zero applied address flux, the switch is closed, and the TES current flows in parallel through its Josephson junctions, which are in a zero-voltage state. At a flux of $\Phi_0/2$, the combined current flow through the parallel Josephson junctions in each switch drops near zero, and the switch acts as a normal resistor with a value orders of magnitude larger than the TES resistance. The `open' resistance is large enough that it introduces no significant Johnson-Nyquist current noise. The switches used in I-CDM consist of four Josephson junctions rather than two (Fig. \ref{fig:CurrentSteering}b), which allows both larger operating margins and higher current-carrying capacity than two-junction switches\cite{Zappe1977}. 
The new switches work well, have high yield, and have wide operating margins.

\section{Current-Summed (I-CDM) Array Architecture}\label{array}
\begin{figure}
\begin{center}
\includegraphics[%
  width=1.0\linewidth,
  keepaspectratio]{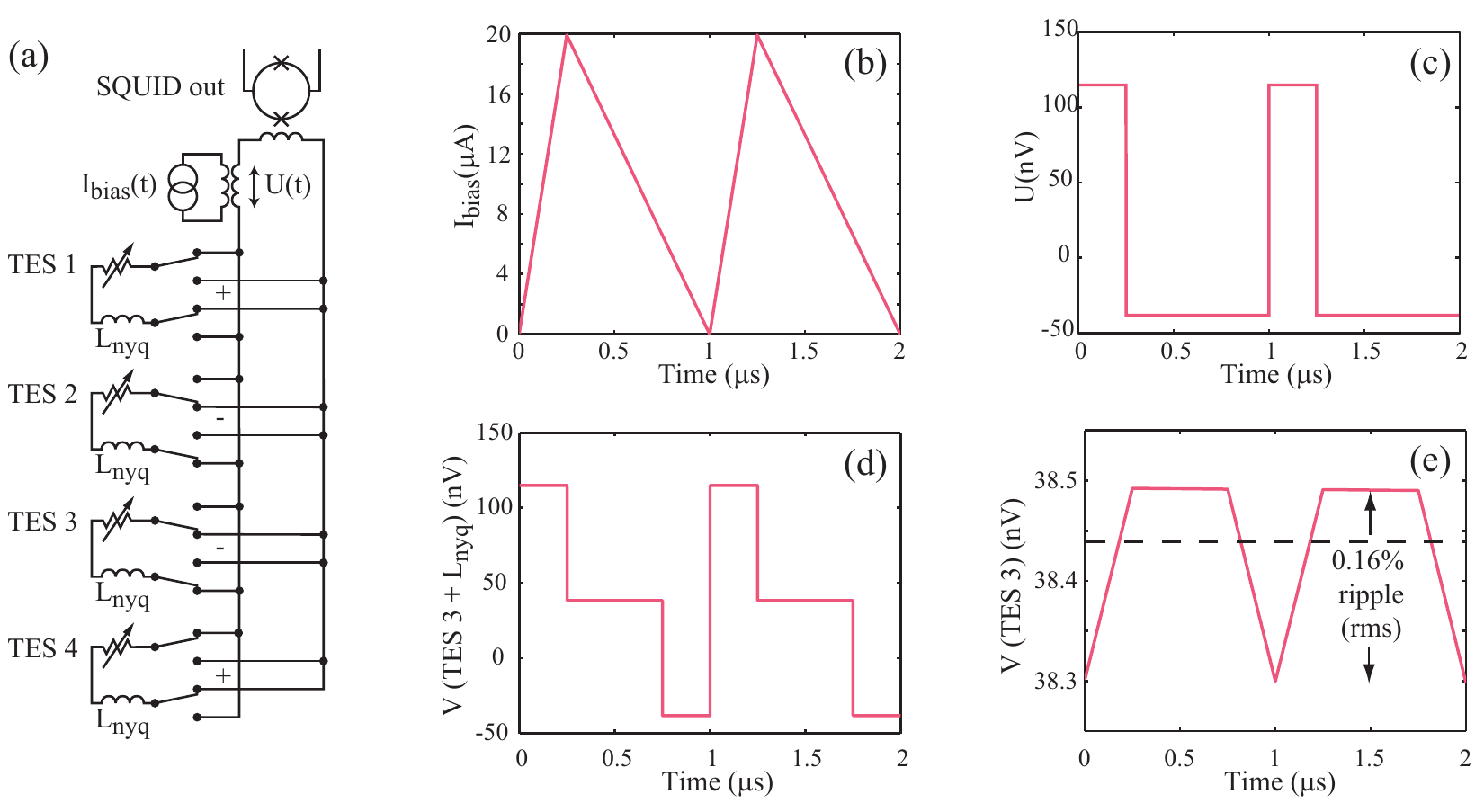}
\end{center}
\caption{
The I-CDM multiplexer. (a) A schematic of a four-pixel I-CDM multiplexer. The currents through all TES pixels (variable resistors in the figure) and their Nyquist inductors $L_{\rm{nyq}}$ are modulated in a Walsh pattern and summed in parallel into the input coil of a SQUID.
In the example state shown in the schematic, TES pixels 1 and 4 are coupled to the SQUID with positive polarity, while TES pixels 2 and 3 are coupled with negative polarity. A voltage bias is applied to the detectors by means of a current $I_{\rm{bias}}$, which is injected into the primary of a coupled inductor. $I_{\rm{bias}}$ induces a voltage $U(t)$ on the secondary of the inductor. (b) A periodic bias current ramp $I_{\rm{bias}}$ chosen to bias TES 1 at zero voltage, and TES 2-4 at $\approx$ 38 nV. (c) The induced voltage levels $U(t)$ on the secondary. (d) The induced voltage $V(t)$ on the series combination of TES 3 and its Nyquist inductor $L_{\rm{nyq}}$
(e) The voltage across TES 3, which is approximately constant except for a 0.16 \% rms ripple. }
\label{fig:ICDMfull}
\end{figure}

The I-CDM array architecture presented here uses the polarity modulators shown in Fig. \ref{fig:CurrentSteering}. 
Figure \ref{fig:ICDMfull}a shows the I-CDM circuit diagram for a small 4-pixel array. In this circuit, each TES is wired in series with a Nyquist inductor that is large enough that the current through the TES is approximately constant during a multiplexed frame. Each TES (and its Nyquist inductor) is coupled to a polarity modulator, schematically represented in the figure as two single-pole double-throw (SPDT) switches that connect the two electrodes of the TES and Nyquist inductor alternately to the two wires coming from the SQUID coil. The current from all four TESs is summed with different polarities into one pair of wires, with the polarity of each summation determined by the state of the associated pair of SPDT switches. The column is read out with a single SQUID amplifier on the same silicon chip. 

In I-CDM, the TES detectors are dc biased. The average voltage bias level $\overline{V}$ on each pixel is set by applying a repeating linear current ramp $I_{\rm{bias}}$(t) to the coupled inductor in Fig. \ref{fig:ICDMfull}a. One example of an $I_{\rm{bias}}$(t) pattern is shown in Fig. \ref{fig:ICDMfull}b. The current ramp $I_{\rm{bias}}$(t) induces a repeating series of voltage levels $U(t)$ on the secondary of the coupled coil (Fig. \ref{fig:ICDMfull}c). The polarity of the coupling between each pixel and the bias signal $U(t)$ is modulated in a Walsh code. Figure \ref{fig:ICDMfull}d shows the modulated voltage bias across the series combination of TES 3 and its Nyquist inductor, $V(\rm{TES3} + L_{\rm{nyq}})$, for pixel 3. In the four-pixel example in Fig. \ref{fig:ICDMfull}, the vector of average voltages $\overline{V}$ seen by the four TES pixels is 
$\overline{V}_i = \sum W_{ij}U_j/4$ (summed over j=1..4), where $W_{ij}$ is the 4$\times$4 Walsh
matrix and $U_j$ is the value of the voltage $U(t)$ for each pixel during the four Walsh periods. The voltage across the TES itself stays approximately constant because the impedance of $L_{\rm{nyq}}$ is large compared to the TES resistance at the modulation frequency. The voltage across pixel 3, V(TES3), is shown as an example in Fig. \ref{fig:ICDMfull}e.

All Walsh matrices are non-singular, thus any combination of average TES bias voltages $\overline{V}$ can be established by multiplying the desired values of $\overline{V}$ by the inverse of the Walsh matrix, $W_{ij}^{-1}=(1/4)W_{ij}$. The first TES is typically disconnected because its output is not modulated in the Walsh code, so it doesn't share the same benefits from nulling common-mode pickup. For large multiplex factors, this results in only a small loss in the number of pixels.
Thus, we set $\overline{V_1}=0$, which has the added benefit that $\rm{I_{bias}}$ will return to the same level after each frame. The four voltages on the secondary $U_i$ must be $U_i= \sum W_{ij}\overline{V}_i$ (summed over j=1..4), or

\vspace{-5 pt}
\begin{equation}
\vspace{-3 pt}
\label{U}
\begin{pmatrix} U_1 \\ U_2 \\ U_3 \\ U_4 \\ \end{pmatrix} = \begin{pmatrix} 1 & 1 & 1 & 1 \\ 1 & 1 & -1 & -1 \\ 1 & -1 & -1 & 1 \\
 1 & -1 & 1 & -1\\
 \end{pmatrix} \begin{pmatrix} 0 \\ \overline{V_2} \\ \overline{V_3} \\ \overline{V_4} \\\end{pmatrix}.
\end{equation}

The example of Fig. \ref{fig:ICDMfull} is for the case in which all of the TES voltages are chosen to have the same value $V_0$. In this case, $U_1=3V_0$, and $U_2=U_3=U_4=-V_0$. The numerical values chosen in Fig. \ref{fig:ICDMfull} are for a particular TES detector design, with $V_0$=38 nV, row periods of 250 ns, and $L_{nyq}$=100~nH. 
In this example, the voltage across the TES is approximately constant (Fig. \ref{fig:ICDMfull}e) with an rms ripple of only 0.16 \%. Since this ripple occurs over periods much shorter than the response time of the TES, it does not measurably degrade the achievable energy resolution.

As the polarity of each TES is switched, it generates a back-action voltage. This would first appear to be a source of crosstalk, since it also acts on other TES pixels. However, over a full frame, the crosstalk back-action is null due to the orthogonality of the Walsh vectors. The back-action of the switching on the TES itself, averaging over multiple frames, appears as 
an additional source of resistance $R_{\rm{s}}=L_{\rm{sw}} / \tau_{\rm{dwell}}$ in series with the voltage bias \cite{Irwin2010}, where $\tau_{\rm{dwell}}$ is the average time between switching. The source resistance, $R_{\rm{s}}$, must be kept small compared to the TES bias resistance $R_0$ to maintain a voltage bias. Another constraint is placed by Josephson-frequency oscillations: the voltage applied to the `open' switch will cause a small ac leakage current to oscillate at the frequency $V/\Phi_0$. The switch circuit must be designed so that the Josephson oscillations are out of band.

\begin{figure}
\begin{center}
\includegraphics[%
  width=0.65\linewidth,
  keepaspectratio]{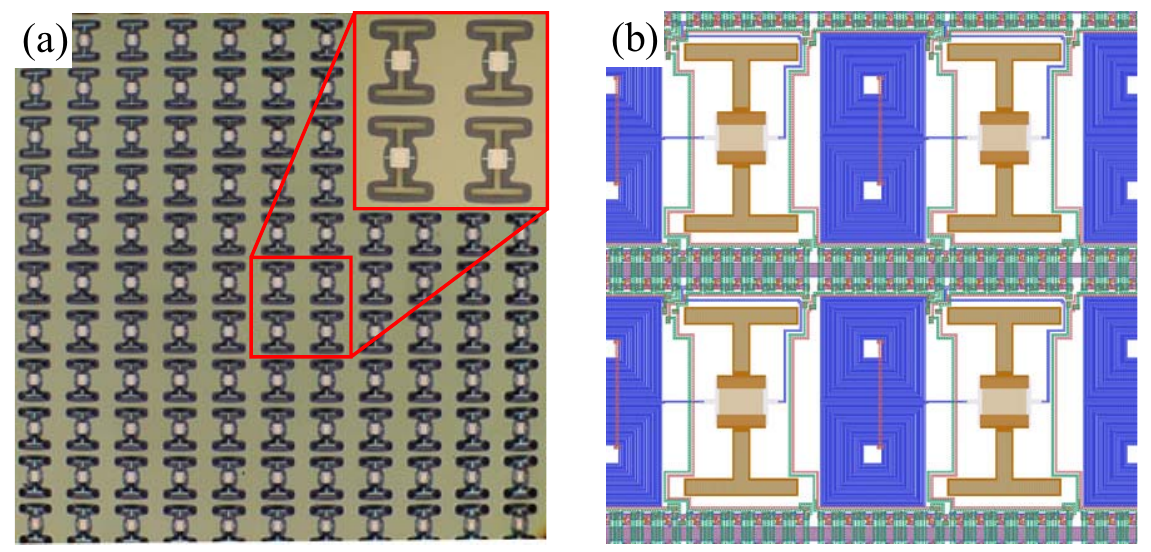}
\end{center}
\caption{
(a) A photograph of part of a 32$\times$32 TES x-ray calorimeter array that was fabricated as a geometric test. This array uses Mo-Cu TES thermometers, and will later be integrated with x-ray absorbers cantilevered over the multiplexer components. Inset: a magnified view of four pixels. The area of relieved silicon nitride membranes is the darker outline around the `I'-shaped Cu x-ray absorber attachment structures. (b) A full lithographic layout of the I-CDM multiplexer, which fits beneath overhanging x-ray absorbers. Each two-lobed blue coil is a Nyquist inductor. The symmetry of the lobes ensures that the magnetic field on the adjacent TESs is close to zero. The polarity switches are the circuit elements running horizontally between the I-shaped posts.}
\label{fig:ICDMArray}
\end{figure}

I-CDM has great potential to increase the scalability of both TES bolometer instruments for far-infrared / millimeter-wave measurements, and for TES x-ray detectors. In order to demonstrate the potential to incorporate I-CDM multiplexer components within an x-ray detector pixel, we have fabricated a 32$\times$32 TES x-ray calorimeter test array with pixels on a 300 $\mu$m pitch, and room for the I-CDM multiplexer components (Fig. \ref{fig:ICDMArray}a). In a full implementation, x-ray absorbers will be cantilevered over the multiplexer components, and connected to the `I'-shaped absorber attachment structures. We have also developed a full lithographic layout of an I-CDM multiplexer integrated in this test array (Fig. \ref{fig:ICDMArray}b).

\section{Conclusions}
I-CDM uses compact multiplexing elements that fit beneath an x-ray absorber in a TES array with a 300 $\mu$m pixel pitch. I-CDM modulation elements are much smaller than the LC filters used in FDM and the microwave resonators used in MKIDs and microwave SQUIDs. I-CDM does not use shunt resistors to voltage bias TES detectors, greatly reducing the power dissipation, and making it possible to scale to larger arrays. 

The output bandwidth provided by dc SQUID amplifiers is typically a few megahertz, which limits the total multiplexing factor. Greater output bandwidth and much higher multiplexing factors can be achieved by using microwave SQUID multiplexers\cite{Mates2008} as the readout SQUIDs in I-CDM instead of traditional dc SQUIDs.

\vspace{-10pt}
\begin{acknowledgements}
We acknowledge support from NASA under grant NNG09WF27I.
\end{acknowledgements}


\begin{thebibliography}{99}
\bibitem{Irwin1995}
K.D. Irwin, {\it Appl. Phys. Lett.} \textbf{66}, 1998, (1995).

\bibitem{Chervenak1999}
J.A. Chervenak, K.D. Irwin, E.N. Grossman, J.M. Martinis, C.D. Reintsema, and M.E. Huber, {\it Appl. Phys. Lett.} \textbf{74}, 4043, (1999).

\bibitem{Yoon2001}
J. Yoon, J. Clarke, J.M. Gildemeister, A.T. Lee, M.J. Myers, P.L. Richards, and J.T. Skidmore, {\it Appl. Phys. Lett.} \textbf{78}, 371, (2001).

\bibitem{Karasik2001}
B. Karasik, and W. McGrath, {\it Proc. of 12th Int'l Symp. on Space Terahertz Technology}, 436, (2001).

\bibitem{Podt2002}
M. Podt, J. Weenink, J. Flokstra, and H. Rogalla, {\it Physica C} \textbf{368}, 218, (2002).

\bibitem{Irwin2010}
K.D. Irwin, M.D. Niemack, J. Beyer, H.M. Cho, W.B. Doriese, G.C. Hilton, C.D. Reintsema, D.R. Schmidt, J.N. Ullom, and L.R. Vale, {\it Supercond. Sci. Technol.} \textbf{23}, 034004, (2010).

\bibitem{Gottardi2009}
L. Gottardi, J. van der Kuur, P.A.J. de Korte, R. Den Hartog, B. Dirks, M. Popescu, and H.F.C. Hoevers, {\it AIP Conference Proceedings} \textbf{1185}, 538, (2009).

\bibitem{Fowler2011}
J.W. Fowler, W.B. Doriese, G.C. Hilton, K.D. Irwin, D.R. Schmidt, G. Stiehl, J.N. Ullom, and L.R. Vale, {\it Proc. of 14th Int'l Workshop on Low Temp. Detectors, Heidelberg, Germany, Aug. 1-5, 2011}, submitted.

\bibitem{Niemack2010}
M.D. Niemack, J. Beyer, H.M. Cho,  W.B. Doriese, G.C. Hilton, K.D. Irwin, C.D. Reintsema, D.R. Schmidt, J.N. Ullom, and L.R. Vale, {\it Appl. Phys. Lett.} \textbf{96}, 1635093, (2010).

\bibitem{Zappe1977}
H.H. Zappe, {\it IEEE Trans. on Magnetics} \textbf{13}, 41, (1977).

\bibitem{Beyer2008}
J. Beyer, and D. Drung, {\it Supercond. Sci. Technol.} \textbf{21}, 105022, (2008).

\bibitem{Mates2008}
J.A.B. Mates, G.C. Hilton, K.D. Irwin, L.R. Vale, and K.W. Lehnert, {\it Appl. Phys. Lett.} \textbf{92}, 023514, (2008).




\end{thebibliography}
\end{document}